\documentclass[preprint,aps,prd,showpacs,superscriptaddress,nofootinbib,tightenlines]{revtex4}

\usepackage{amsfonts}
\usepackage{multirow}
\usepackage{mathrsfs}
\usepackage{graphicx}
\usepackage{amsmath}
\usepackage{amssymb}
\usepackage{bm}
\usepackage{bbm}
\usepackage{color}
\usepackage{array}

\def\OMIT#1{}

\newcommand{\nn}{\nonumber}

\newcommand{\beq}{\begin{equation}}
\newcommand{\eeq}{\end{equation}}
\newcommand{\bqa}{\begin{eqnarray}}
\newcommand{\eqa}{\end{eqnarray}}

\begin{document}
\title{Gluon fragmentation into $B_c^{(*)}$ in NRQCD factorization
}
\author{Feng Feng\footnote{Email: F.Feng@outlook.com}}
\affiliation{China University of Mining and Technology, Beijing 100083, China\vspace{0.2cm}}
\affiliation{Institute of High Energy Physics, Chinese Academy of Sciences, Beijing 100049, China\vspace{0.2cm}}

\author{Yu Jia\footnote{E-mail: jiay@ihep.ac.cn}}
\affiliation{Institute of High Energy Physics, Chinese Academy of Sciences, Beijing 100049, China\vspace{0.2cm}}
\affiliation{School of Physical Sciences, University of
Chinese Academy of Sciences, Beijing 100049, China\vspace{0.2cm}}

\author{Deshan Yang\footnote{E-mail: yangds@ucas.ac.cn}}
\affiliation{School of Physical Sciences, University of
Chinese Academy of Sciences, Beijing 100049, China\vspace{0.2cm}}
\affiliation{Institute of High Energy Physics, Chinese Academy of Sciences, Beijing 100049, China\vspace{0.2cm}}

\date{\today}
\begin{abstract}
{
The universal fragmentation functions of gluon into the flavored quarkonia $B_c$ and (polarized) $B_c^*$ are computed within
NRQCD factorization framework, at the lowest order in velocity expansion and strong coupling constant.
It is mandatory to invoke the DGLAP renormalization program to render the NRQCD short-distance coefficients UV finite in
a point-wise manner. The calculation is facilitated with the sector decomposition method, with the final results
presented with high numerical accuracy.
This knowledge is useful to enrich our understanding toward the
large-$p_T$ behavior of $B_c^{(*)}$ production at LHC experiment.
}

\end{abstract}

\maketitle

Like parton distribution functions (PDFs), fragmentation functions (FFs) are process-independent functions that
encapsulate nonperturbative hadronization effect, and play central role in QCD phenomenology of collider physics.
In accordance with the celebrated QCD factorization theorem~\cite{Collins:1989gx},
the inclusive production rate of an identified hadron $H$ at large transverse momentum in high energy collision
experiment is dictated by the fragmentation mechanism:
\beq
d\sigma[A+B\to H(p_T)+X]  = \sum_i d{\hat\sigma}[A+B\to i(p_T/z)+X] \otimes
D_{i \to H}(z,\mu)+{\mathcal O}(1/p_T^2),
\label{QCD:factorization:theorem}
\eeq
where $A$, $B$ represent two colliding particles, $d\hat{\sigma}$ denotes the inclusive rate for producing the parton $i$,
$D_{i \to H}(z)$ is the fragmentation function for the parton $i$ into $H$,
which characterizes the probability for $i$ to transition into any final state containing the hadron $H$ specifying
the fractional light-cone momentum $z$ with respect to the parent parton $i$.
The sum in (\ref{QCD:factorization:theorem}) is extended over all species of partons, $i=q,\bar{q},g$.
Similar to the PDFs, the scale dependence of fragmentation functions is also
governed by the celebrated Dokshitzer-Gribov-Lipatov-Altarelli-Parisi (DGLAP) equation. Taking the
gluon fragmentation into the hadron $H$ as example, the DGLAP equation reads
\beq
{d \over d \ln \mu^2} D_{g \to H}(z,\mu) =  \sum_{i} \int_z^1 {d\xi\over \xi}
P_{i g} (\xi, \alpha_s(\mu)) D_{i \to H}\left({z\over \xi},\mu\right),
\label{DGLAP:evolution}
\eeq
where $\mu$ is interpreted as the renormalization scale, and $P_{ig}(\xi)$ are the corresponding
splitting kernels. Once this FF is deduced at some initial scale $\mu_0$ by any means,
one then deduce its form at any other scale $\mu$ by solving the evolution equation
(\ref{DGLAP:evolution}).

The fragmentation functions for light hadrons such as $\pi$, $K$, $p$, \ldots,
are hopelessly nonperturbative objects, which can only be extracted from experimental data.
On the contrary, it was realized in the mid-90s that FFs for heavy quarkonia
need not be genuinely nonpertubative entities, which nevertheless can be largely understood in
perturbative QCD by the virtue of asymptotic freedom~\cite{Braaten:1993mp,Braaten:1993rw}.
This philosophy is systematically embodied in the modern nonrelativistic QCD (NRQCD) factorization framework~\cite{Bodwin:1994jh},
that is, the quarkonium FFs can be expressed as the sum of products of short-distance coefficients (SDCs) and long-distance yet universal
NRQCD matrix elements, with the series organized by the velocity expansion~\cite{Braaten:1993mp,Braaten:1993rw}.
Since the nonperturbative NRQCD matrix elements are merely numbers rather than functions, the profiles of the quarkonia FFs
are largely determined by perturbation theory, therefore NRQCD factorization approach is endowed with
strong predictive power.

During the past three decades, a number of fragmentation functions for $S$-wave/$P$-wave charmonia/bottomonia have been investigated in the context of
NRQCD factorization approach, typically at lowest order in $\alpha_s$ (for a very incomplete list, see~\cite{Braaten:1993mp,Braaten:1993rw,Braaten:1993jn,Ma:1994zt,Braaten:1994kd,Cho:1994qp,Ma:1995ci,Ma:1995vi,Braaten:1995cj,Cheung:1995ir,Qiao:1997wb,Braaten:2000pc,
Bodwin:2003wh,Sang:2009zz,Bodwin:2012xc,Ma:2013yla,Artoisenet:2014lpa,Bodwin:2014bia,Gao:2016ihc,Sepahvand:2017gup,Zhang:2017xoj,Feng:2017cjk,Yang:2019gga}.
With the advance of higher-order calculational technique, the SDCs associated with some $S$-wave quarkonium FFs have recently
been calculated through the next-to-leading order (NLO) in $\alpha_s$~\cite{Braaten:2000pc,Artoisenet:2014lpa,Artoisenet:2018dbs, Feng:2021uct, Zhang:2018mlo, Feng:2018ulg}.

Unlike $J/\psi$ and $\Upsilon$, the $B_c$ meson is the unique heavy quarkonium which are composed of two different heavy flavors:
the $b$ and $\bar{c}$ quarks. It is interesting to understand the production mechanism of this special heavy meson in hadron collision environment.
The LO fragmentation functions for $b/\bar{c}\to B_c^{(*)}$ was computed long ago. The NLO perturbative correction
has also been recently available. Due to the rich gluon content inside the proton in small $x$ region, it is also of great
phenomenological incentive to study the $g\to B_c^{(*)}$ fragmentation functions to predict their production rates
at \texttt{LHC}. To produce a $B_c$ meson in this case, one has to first create a pair of $c\bar{c}$ and another pair of $b\bar{b}$,
therefore the perturbative order of this LO fragmentation process is already comparable with that of the
NLO QCD correction to $b/\bar{c}\to B_c^{(*)}$. Thus, the computation of the  $g\to B_c^{(*)}$ fragmentation functions at LO is already
rather challenging technically, which have never been considered before.
The aim of this work is to fill this gap, by computing the fragmentation functions for
$g\to B_c^{(*)}$ at lowest order in velocity expansion and $\alpha_s$, by invoking the sector decomposition technique
widely used in the area of multi-loop computation.

In literature there are several different strategies to extract the quarkonium FFs. Among them,
the most systematic approach is to start from the gauge-invariant operator definition for the fragmentation functions pioneered
by Collins and Soper long ago~\cite{Collins:1981uw} (Note that this definition was first used by Ma to compute the quarkonium FFs in NRQCD~\cite{Ma:1994zt}).
One great virtue of this operator-based approach is to render renormalization program transparent.
According to the operator definition given in \cite{Collins:1981uw},
the $g$-to-$B^{(*)}_c$ fragmentation function is expressed as (see also \cite{Bodwin:2003wh,Bodwin:2012xc}):
\bqa \label{CS:def:Fragmentation:Function}
& & D_{g \to B^{(*)}_c}(z,\mu) =
\frac{-g_{\mu \nu}z^{D-3} }{ 2\pi k^+ (N_c^2-1)(D-2) }
\int_{-\infty}^{+\infty} \!dx^- \, e^{-i k^+ x^-}
\\
&& \times
\langle 0 | G^{+\mu}_c(0)
\Phi^\dagger(0,0,{\bf 0}_\perp)_{cb}  \sum_{X} |B^{(*)}_c(P)+X\rangle \langle B^{(*)}_c(P)+X|
\Phi(0,x^-,{\bf 0}_\perp)_{ba} G^{+\nu}_a(0,x^-,{\bf 0}_\perp) \vert 0 \rangle.
\nn
\eqa
$D=4-2\varepsilon$ signifies the space-time dimensions, and $\mu$ is the renormalization scale associated with
this nonlocal composite operator. $G_{\mu\nu}$ is the matrix-valued gluon field-strength tensor in the adjoint representation of $SU(N_c)$.
Here it is convenient to adopt the light-cone coordinates.
Any four-vector $A^\mu=(A^0,A^1,A^2,A^3)$ can be recast in the light-cone format $A^\mu=(A^+,A^-, {\bf A}_\perp)$,
with $A^\pm \equiv {1\over \sqrt{2}}(A^0\pm A^3)$ and ${\bf A}_\perp \equiv (A^1,A^2)$.
We assume to work in a frame where the $B_c$ meson is moving along the $z$ direction with the $+$-momentum $P^+$,
Moreover, $k^+ = P^+/z$ denotes the $+$-component momentum of injected by the gluon field strength operator.
The symbol $X$ indicates collectively those unobserved light hadrons accompanying the $B_c$.

The eikonal factor $\Phi(0,x^-,{\bf 0}_\perp)$ in (\ref{CS:def:Fragmentation:Function}) is a
path-ordered exponential of the gluon field,  inserted to ensure the gauge invariance of the FF:
\beq
\Phi(0,x^-,{\bf 0}_\perp)_{ba} = \texttt{P} \exp
\left[ i g_s \int_{x^-}^\infty d y^- A^+(0^+,y^-,{\bf 0}_\perp) \right]_{ba},
\label{Gauge:Link:Definition}
\eeq
where $A^\mu$ designates the matrix-valued gluon field in the adjoint representation,
$g_s$ is the QCD coupling constant, and $\texttt{P}$ denotes the path-ordering,

The key observation underlying the NRQCD factorization that, prior to forming a $B_c$ state via soft interaction,
the $b$ and $\bar{c}$ quarks have to be first created with slow relative momentum, also within small distance $\sim 1/m_{c,b}$,
which necessarily involves hard momentum transfer, thus can be studied in perturbation theory owing o asymptotic freedom of QCD.
At the lowest order in velocity expansion, NRQCD factorization~\cite{Bodwin:1994jh} allows
one to refactorize the $g\to B_c^{(*)}$  FFs into the product of the short-distance coefficients and the long-distance NRQCD matrix elements:
\begin{subequations}
\label{FF:NRQCD:factorization}
\bqa
D_{g \rightarrow B_{c}}(z, \mu)&=&d_{g\to B_{c}}(z, \mu)\frac{2\pi}{N_c}\frac{\left\langle 0\left|\mathcal{O}_{1}^{B_{c}}\left({ }^{1} S_{0}\right)\right| 0\right\rangle}{ M^{3}}+\cdots, \\
\quad D_{g \rightarrow B_{c}^{*}}(z, \mu)&=&d_{g\to B_{c}^{*}}(z, \mu)\frac{1}{D-1}\frac{2\pi}{N_c}\frac{\left\langle 0\left|\mathcal{O}_{1}^{B_{c}^{*}}\left({ }^{3} S_{1}\right)\right| 0\right\rangle}{M^{3}}+\cdots\,,
\eqa
\end{subequations}
where $M=m_b+m_c$, $D$ is the dimension of space-time,
$d_{g\to H}(z,\mu)$ is the desired SDCs for $H=B_c,B_c^*$, and the corresponding NRQCD production operators are defined by
\begin{subequations}
\label{NRQCD:production:operators}
\bqa
\mathcal{O}_1^{B_c}(^1S_0)  &=&
\chi_c^\dagger \psi_b  \sum_{X} |B_c+X\rangle  \langle B_c+X|
\psi_b^\dagger \chi_c, \\
\mathcal{O}_1^{B^*_c}(^3S_1)  &=&
\chi_c^\dagger \bm{\sigma} \psi_b \sum_{X} |B^*_c+X\rangle \cdot \langle B^*_c+X|
\psi_b^\dagger \bm{\sigma} \chi_c,
\eqa
\end{subequations}
where $\psi_b$ and $\chi_c^\dagger$ are the NRQCD field operators
that annihilate a $b$ quark and a $\bar{c}$ quark, respectively.

Using the vacuum saturation approximation, these vacuum NRQCD matrix elements can be appxomated by the radial wave function at the origin
for $B_c^{(*)}$ in phenomenological potential model:
\beq
 \langle {\cal O}_1^{B_{c}}(^1S_0) \rangle \simeq \frac{N_{c}}{2\pi} \vert R(0) \vert^{2},\qquad
 \langle {\cal O}_1^{B_{c}}(^3S_1) \rangle \simeq (D-1)\frac{N_{c}}{2\pi} \vert R(0) \vert^{2}.
\eeq

We can proceed to calculate the SDCs $d^{H}_{1}(z)$ by the standard perturbative matching technique, {\it i.e.},
by replacing the physical $H$ state in (\ref{FF:NRQCD:factorization}) with the free quark pair $b\bar{c}$ carrying the appropriate quantum number.
Concretely speaking, one replaces $B_c$ by a fictitious meson $b\bar{c}(^1S_0^{(1)})$, and replaces
$B_c^*$ by the quark state $b\bar{c}(^3S_1^{(1)})$. Computing the left side of (\ref{FF:NRQCD:factorization}) using perturbative QCD, combined
with the following vacuum matrix elements in perturbative NRQCD:
\begin{equation}
 \langle {\cal O}_1^{b \bar{c}}(^1S_0) = 2N_c, \qquad  \langle {\cal O}_1^{b \bar{c}}(^3S_1) = 2N_c(D-1),
\end{equation}
we can directly solve the SDC $d^H_{1}(z)$, order by order in $\alpha_s$.

\begin{figure}[tbh]
\centering
\includegraphics[width=0.45\textwidth]{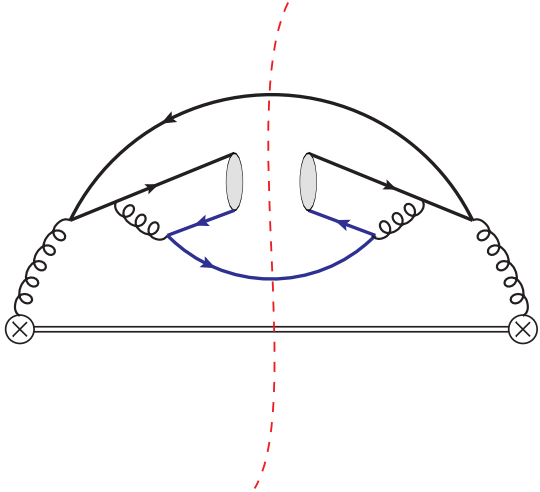}
\caption{Representative cut Feynman diagram for perturbative gluon fragmentation function $D_{g\to b\bar{c}}(z)$ at LO
in $\alpha_s$. The cap represents the gluonic field strength operator $G_a^{+\nu}$, and double line signifies the eikonal line.
\label{Feynman:diagram}}
\end{figure}

The fragmentation function defined in (\ref{CS:def:Fragmentation:Function}) is manifestly gauge-invariant. In practical calculation,
we specialize to Feynman gauge for simplicity.
We use a private \texttt{Mathematica} code to automatically generate the Feynman diagrams and the associated
cut amplitudes that correspond to the perturbative fragmentation function defined in \eqref{CS:def:Fragmentation:Function}.
The Feynman rules for the eikonal propagator and vertex~\cite{Collins:1981uw}, as well as those for conventional QCD propagators and vertices
are also implemented with the aid of the package \texttt{Qgraf}~\cite{Nogueira:1991ex}.
There are seven diagrams on each side of the cut, so in total 49 cut diagrams at lowest order in $\alpha_s$.
A simplifying feature is that the gluon propagator cannot be attached to the eikonal line at this perturbative order.
For concreteness, in Fig.~\ref{Feynman:diagram} we exhibit a typical LO cut diagram associated with
the perturbative FF $D_{g\to b\bar{c}}$.

The cut-amplitude structure of fragmentation function stemps from the insertion of the asymptotic out states
in (\ref{CS:def:Fragmentation:Function}). Consequently,
the corresponding cut-line phase space integration measure reads~\cite{Bodwin:2003wh,Bodwin:2012xc}
\bqa
d\Phi_n &=&  4\pi M_{B_c} \delta(k^+-P^+-\sum_{i=1}^n k_i^+) \prod_{i=1}^2 \frac{dk^+_i}{2k_i^+}\frac{d^{D-2}k_{i\perp}}{(2\pi)^{D-1}} \theta(k^+_i),\label{phase:space}
\label{phase:space:integration:measure}
\eqa
where $k_i$ ($i=1,2$) stands for the momentum of the $i$-th on-shell quark line ($\bar{b}$ and $c$)  that pass through the cut.
The integration over $k_i^+$ can be transformed into
a parametric integration in a finite interval, but the integration over the transverse momentum
$k_{i,\perp}$ are completely unbounded, {\it i.e.}, from $-\infty$ to $+\infty$.
This feature may persuade us that integration over $k_{i,\perp}$ could be regarded as loop integration in $D-2$-dimensional spacetime,
with $D=4-2\varepsilon$. Throughout this work, we adopt the dimensional regularization to regularize occurring UV divergences.

To project the $b\bar{c}$ pair onto the intended color/spin/orbital/color states, it is convenient to employ the
covariant projector technique to expedite the calculation. At the lowest order in velocity expansion, it is
legimate to partition the quark momenta inside the fictitious $B_c^{(*)}$ state commensurate to their mass ratio:
$p_{\bar c}=r P$, $p_b= \bar{r} P$, with $r\equiv m_c/(m_b+m_c)$ and $\bar{r}\equiv m_b/(m_b+m_c)$.
One then make the following substitutions in the quark amplitude to project out the desired contributions from
the $b\bar{c}(^1S_0^{(1)})$ and $b\bar{c}(^3S_1^{(1)})$~\cite{Petrelli:1997ge}:
\begin{subequations}
\bqa
v(p_{\bar c}) \bar{u}(p_b) &\longrightarrow & {1\over \sqrt{8 m_b m_c}} \left({p\!\!\!\slash_{\bar c}}-m_c\right) \gamma_5
\left({p\!\!\!\slash_b}+m_b\right) \otimes { \texttt{1}_c  \over \sqrt{N_c}},\quad {\rm for}\;{}^1S_0
\\
v(p_{\bar c}) \bar{u}(p_b) &\longrightarrow & {1\over \sqrt{8 m_b m_c}} \left({p\!\!\!\slash_{\bar c}}-m_c\right) \varepsilon^*\!\!\!\!\!\slash
\left({p\!\!\!\slash_b}+m_b\right) \otimes { \texttt{1}_c  \over \sqrt{N_c}},\quad {\rm for}\;{}^3S_1
\eqa
\label{spin:color:projectors}
\end{subequations}
with $p_b^2=m_b^2$, $p_{\bar c}^2=m_c^2$, and $\varepsilon^\mu(P)$ designates the polarization vector for the fictitious
$B_c^*$ meson.

With the aid of the above projectors \eqref{spin:color:projectors}, we employ the packages \textsf{FeynCalc/FormLink}~\cite{Mertig:1990an,Feng:2012tk}
to conduct the Dirac/color trace operation. We also use the package \textsf{Apart}~\cite{Feng:2012iq} to simplify the
amplitude by the method of partial fraction, to make the integrand simpler.

It is appealing to use some modern multi-loop technique such as reverse unitarity method and integration by parts (IBP) to
deal with phase space integration in \eqref{phase:space:integration:measure}. The gluon fragmentation into
$c\bar{c}({}^3S_1^{(1)})$ and $c\bar{c}({}^1P_1^{(1)})$ have been analytically
calculated in this manner~\cite{Zhang:2017xoj,Feng:2017cjk,Zhang:2018mlo}. Nevertheless,
it turns out to be a quite subtle issue to use these techniques to deal with our case.
In this work, similar to \cite{Feng:2018ulg,Feng:2021uct}, we choose to apply the {\it sector decomposition} method~\cite{Binoth:2000ps,Binoth:2003ak} to evaluate
the phase space integration as given in \eqref{phase:space:integration:measure}.
Consequently, we will present our final results in an entirely numerical manner. In our opinion, the approach adopted in this work
appears to be more amenable to automated calculation, and yield more accurate numerical predictions
than the complicated subtraction approach first developted in \cite{Artoisenet:2014lpa} (see also \cite{Artoisenet:2018dbs,Zheng:2019gnb,Zheng:2019dfk}).

We first combine all the propagators in a cut diagram using Feynman parametrization, then
accomplish two-loop integration over $k_{1,2\:\perp}$ in $D-2$-dimensional spacetime.
We are then left with multi-fold integrals over Feynman parameters, which can be numerically calculated by the package
\textsf{FIESTA}~\cite{Smirnov:2013eza} which is based on the sector decomposition algorithm~\cite{Binoth:2000ps,Binoth:2003ak}.
This method is typically useful with many finite multi-variable parametric integrals as output,
with various UV poles explicitly isolated.

Upon summing all 49 cut diagrams, we find the total cut amplitude still contains a single uncanceled pole,
whose coefficient varies with the momentum fraction $z$. This pole is clearly of UV origin, from the large
$k_{1,2\:\perp}$ integration region in \eqref{phase:space:integration:measure}.
This is a clear sign that the bare fragmentation function requires an additional
operator renormalization, following the DGLAP paradigm~\cite{Collins:1981uw,Bodwin:2014bia}:
\beq
D^{\overline{\rm MS}}_{g\to H}(z,\mu) = D_{g\to H}(z,\mu) - \sum_{q=b,\bar{c}}
{1\over \epsilon}  \int_z^1 \!\!{dy\over y}\, P_{qg}(y) D_{q\to H}(z/y,\mu),
\label{DGLAP:renormalization}
\eeq
where $P_{qg}(y)$ represents the splitting kernel for $g\to q$:
\beq
P_{qg}(y) = {\alpha_s(\mu)\over 2\pi} T_F\left[ y^2+(1-y)^2 \right]+{\cal O}(\alpha_s^2),
\eeq
with $T_F=1/2$.

The heavy-quark-to-$B_c^{(*)}$ fragmentation functions at LO in $\alpha_s$ are known long ago~\cite{Braaten:1993jn}.
Recasting the results in the NRQCD factorization language, we have
\begin{subequations}
\begin{eqnarray}
	D_{b \rightarrow B_{c}}(z, \mu)&=&d_{b\to B_{c}}(z, \mu)\frac{2\pi}{N_c}\frac{\left\langle 0\left|\mathcal{O}_{1}^{B_{c}}\left({ }^{1} S_{0}\right)\right| 0\right\rangle}{ M^{3}}+\cdots, \\
	D_{b \rightarrow B_{c}^{*}}(z, \mu)&=&d_{b\to B_{c}^{*}}(z, \mu)\frac{1}{D-1}\frac{2\pi}{N_c}\frac{\left\langle 0\left|\mathcal{O}_{1}^{B_{c}^{*}}\left({ }^{3} S_{1}\right)\right| 0\right\rangle}{M^{3}}+\cdots, \\
	D_{b \rightarrow B_{c}^{*T}}(z, \mu)&=&d_{b\to B_{c}^{*T}}(z, \mu)\frac{1}{D-1}\frac{2\pi}{N_c}\frac{\left\langle 0\left|\mathcal{O}_{1}^{B_{c}^{*}}\left({ }^{3} S_{1}\right)\right| 0\right\rangle}{M^{3}}+\cdots\,.
\end{eqnarray}
\label{Heavy:quark:frag:function}
\end{subequations}
where the second equation implies the polarization-summed $B_c^*$, and the last one implies the
transversely-polarized $B_c^*$ (denoted by $B_c^{*T}$ henceforth). For the $\bar{c}$ fragmentation into
$B_c^{(*)}$, one simply makes the exchange $r\leftrightarrow \bar{r}$ in above expressions.

The corresponding short-distance coefficients in \eqref{Heavy:quark:frag:function} at LO in $\alpha_s$ read~\cite{Yang:2019gga}:
\begin{subequations}
\begin{eqnarray}
 &&d_{b\to B_c}(z)=\frac{\alpha_s^2 C_F^2}{24\pi N_c}\frac{z(1-z)^2}{r^2(1-\bar r z)^6} \\
&&\times \left[6-18(1-2 r)z+(21-74 r+68 r^2)z^2-2\bar r (6-19r+18r^2)z^3+3\bar r ^2(1-2r+2r^2)z^4\right],\nonumber\\
&&d_{b\to B_{c}^{*}}(z, \mu)=\frac{\alpha_s^2 C_F^2}{8\pi N_c}\frac{z(1-z)^2}{\bar r^2(1- r z)^6} \\
&&\times \left[2-2(3-2r)z +3(3-2r+4r^2)z^2    -2\bar r(4-r+2r^2)z^3 +\bar r^{2}(3-2r+2r^2)z^4\right],\nonumber
\\
 &&d_{b\to B_{c}^{*T}}(z, \mu)=\frac{\alpha_s^2 C_F^2}{8\pi N_c}\frac{z(1-z)^2}{\bar r^2(1- r z)^6} \\
&&~~~~~~~~~~~~~~~~~~~~\times \left[2 - 2 (3 - 2 r) z + (9 - 4 r + 10 r^2) z^2 - 2 \bar r (4 + r) z^3 +
 3 \bar r^2 z^4\right].\nonumber
\end{eqnarray}
\end{subequations}

For our purpose, we need compute these LO quark fragmentation functions to $\cal{O}(\epsilon)$ in \eqref{DGLAP:renormalization}.
Note in \eqref{DGLAP:renormalization} the UV pole is subtracted in accordance with the $\overline{\rm MS}$ procedure.

After implementing the DGLAP renormalization program,  we then obtain the pointwise UV-finite SDCs for the
$g$-to-$B_c^{(*)}$ FFs (defined in the $\overline{\rm MS}$ scheme).
It is convenient to divide them into two parts:
\beq
d_{g\to H}(z,\mu) = c^H_0(z) \ln\frac{\mu^2}{M^2}  + \alpha_s^3 c^H_1(z),
\label{Dividing:SDC:two:parts}
\eeq
with the logarithmic part is given by
\bqa
\label{c0:H:def}
c^H_0(z) & = & \sum_{q=b,\bar{c}} \int_{z}^1 \frac{dy}{y} P_{qg}(y) \; d_{q\to H}(z/y).
\eqa

The logarithmic coefficient function of $c^H_0(z)$ can be obtained in closed form. Here we consider three cases,
with $H=B_c$, $B_c^*$, and $B_c^{*T}$:
\begin{eqnarray}
c_0^{B_c}(z)
&=&\frac{\alpha_s^3 C_F^2 }{96 \pi ^2 N_c r^2}\Bigg\{12 z(1-z) \ln z
\nonumber\\
&&
-\frac{2}{\bar r^5}\Bigg[ 2r(5 - 6 r +
 3 r^2) +3 \bar r(1 - 6 r + 10 r^2 - 8 r^3 + 2 r^4) z -6\bar r^5 z^2\Bigg]\ln\frac{1-\bar r z}{r}\nonumber\\
&&+\frac{1}{15 \bar r^5 (1 - \bar r z )^5 r}\Bigg[4 (6 + 40 r^2 - 105 r^3 + 90 r^4 - 31 r^5) \nonumber\\
&&- \bar r(168 - 255 r +
    1760 r^2 - 3255 r^3 + 2460 r^4 - 758 r^5) z \nonumber\\
&&+ \bar r^2 (528 -
    1650 r + 6925 r^2 - 9870 r^3 + 6375 r^4 - 1678 r^5) z^2\nonumber\\
&&-
 5 \bar r^3 (192 - 876 r + 2759 r^2 - 3123 r^3 + 1659 r^4 -
    342 r^5) z^3 \nonumber\\
&&+
 5 \bar r^4 (216 - 1224 r + 3109 r^2 - 2814 r^3 + 1182 r^4 -
    174 r^5) z^4 \nonumber\\
&&- \bar r^5 (744 - 4755 r + 10075 r^2 - 7200 r^3 +
    2250 r^4 - 180 r^5) z^5 \nonumber\\
&&+
 6 \bar r^6 (48 - 325 r + 585 r^2 - 315 r^3 + 60 r^4) z^6 \nonumber\\
&&-
 6 \bar r^7 (8 - 55 r + 85 r^2 - 30 r^3) z^7\Bigg]\Bigg\}+\Bigg(r\leftrightarrow \bar r\Bigg )\nonumber \\
 &\stackrel{z\to 0}{\longrightarrow}&\frac{\alpha_s^3 C_F^2}{360 \pi ^2 N_c \bar r^5 r^3}\Big[\bar r(6 + 6 r + 46 r^2 - 59 r^3 + 31 r^4) +
   15 r^2 (5 - 6 r + 3 r^2)\ln r\Big]\nonumber\\&&+\Bigg(r\leftrightarrow \bar r\Bigg )\,,
\end{eqnarray}
and
\begin{eqnarray}
c_0^{B_c^*}(z)
&=&\frac{\alpha_s^3 C_F^2 }{96 \pi ^2 N_c r^2}\Bigg\{12 (z(1-z)+4 r z^2) \ln z \nonumber\\
&&
-\frac{6}{\bar r^5}\Bigg[ 2 + 2 r^3 - \bar r (1 + 6 r - 10 r^2 + 8 r^3 - 2 r^4) z -
 2 \bar r^5 (1 - 4 r) z^2\Bigg]\ln\frac{1-\bar r z}{r}\nonumber\\
&&+\frac{1}{5 \bar r^4 (1 - \bar r z )^5 r}\Bigg[2 (12 + 17 r + 2 r^2 + 17 r^3 + 12 r^4)\nonumber\\
&&-(168 - 65 r + 370 r^2 - 585 r^3 + 470 r^4 - 238 r^5) z \nonumber\\
&&+\bar r (528 - 710 r + 3045 r^2 - 4440 r^3 + 3245 r^4 - 1038 r^5)  z^2 \nonumber\\
&&- 5 \bar r^2 (192 - 436 r + 1695 r^2 - 2373 r^3 + 1605 r^4 - 414 r^5)   z^3 \nonumber\\
&&+5 \bar r^3 (216 - 654 r + 2383 r^2 - 3186 r^3 + 1970 r^4 - 434 r^5) z^4 z^4 \nonumber\\
&&-\bar r^4 (744 - 2645 r + 9165 r^2 - 11580 r^3 + 6390 r^4 - 1140 r^5) z^5 \nonumber\\
&&+6 \bar r^5 (48 - 185 r + 615 r^2 - 725 r^3 + 340 r^4 - 40 r^5) z^6\nonumber\\
&&-2 \bar r^6 (24 - 95 r + 305 r^2 - 330 r^3 + 120 r^4) z^7\Bigg]\Bigg\}+\Bigg(r\leftrightarrow \bar r\Bigg )\nonumber \\
 &\stackrel{z\to 0}{\longrightarrow}&\frac{\alpha_s^3 C_F^2 }{240 \pi ^2 N_c \bar r^5 r^3}\Big[\bar r (12 + 17 r + 2 r^2 + 17 r^3 + 12 r^4)+30r (1 + r^3) \ln r\Big]\nonumber\\&&+\Bigg(r\leftrightarrow \bar r\Bigg )\,,\\
c_0^{B_c^{*T}}(z)
&=&\frac{\alpha_s^3 C_F^2 }{96 \pi ^2 N_c r^2}\Bigg\{8 z (1 - (1 - 4 r) z)\ln z \nonumber\\
&&
-\frac{4}{\bar r^5}\Bigg[ 2 (1 + r) - \bar r (1 + 8 r - 12 r^2 + 8 r^3 - 2 r^4) z -
 2 \bar r^5 (1 - 4 r) z^2\Bigg]\ln\frac{1-\bar r z}{r}\nonumber\\
&&+\frac{2}{15 \bar r^4 (1 - \bar r z )^5 r}\Bigg[2 (12 + 22 r + 47 r^2 - 28 r^3 + 7 r^4) \nonumber\\
&&- (168 + 5 r + 840 r^2 -
    1575 r^3 + 880 r^4 - 198 r^5) z \nonumber\\
&&+ \bar r (528 - 490 r +
    4145 r^2 - 6630 r^3 + 4045 r^4 - 968 r^5) z^2 \nonumber\\
&&-
 5 \bar r^2 (192 - 356 r + 1951 r^2 - 2871 r^3 + 1755 r^4 -
    402 r^5) z^3 \nonumber\\
&&+
 5 \bar r^3 (216 - 564 r + 2527 r^2 - 3492 r^3 + 2038 r^4 -
    430 r^5) z^4 \nonumber\\
&&- \bar r^4 (744 - 2335 r + 9275 r^2 - 12060 r^3 +
    6450 r^4 - 1140 r^5) z^5 \nonumber\\
&&+
 6 \bar r^5 (48 - 165 r + 605 r^2 - 735 r^3 + 340 r^4 -
    40 r^5) z^6\nonumber\\
&& -
 2 \bar r^6 (24 - 85 r + 295 r^2 - 330 r^3 + 120 r^4) z^7\Bigg]\Bigg\}+\Bigg(r\leftrightarrow \bar r\Bigg )\nonumber \\
 &\stackrel{z\to 0}{\longrightarrow}&\frac{\alpha_s^3 C_F^2 }{240 \pi ^2 N_c \bar r^5 r^3}\Big[\bar r (12 + 22 r + 47 r^2 - 28 r^3 + 7 r^4) + 30 r(1+ r) \ln r\Big]\nonumber\\&&+\Bigg(r\leftrightarrow \bar r\Bigg )\,.
\end{eqnarray}
Notice all of $c_0^H$ are regular near $z=0$.

It is quite challenging, if possible, from the sector decomposition approach  to deduce the analytic expressions for those non-logarithmic
coefficient functions $c^{H}_{1}(z)$ in \eqref{Dividing:SDC:two:parts}.
We are content to knowing their numerical values to very high precision within relatively short time.
For numerical investigation,  we adopt the following input parameters:
\beq
m_c = 1.5\,\text{GeV},\quad m_b=4.8\,\text{GeV},\quad \alpha_s(M)=0.199, \quad 
\eeq
and $\vert R(0)\vert^2 = 1.642\:{\rm GeV}^3$~\cite{Eichten:1995ch}.

\begin{table}[t]
\caption{\label{c1-table}
Numerical values of non-logarithmic coefficient functions $c^H_1(z)$
introduced in \eqref{Dividing:SDC:two:parts}, for $H=B_c$, $B_c^*$ and $B_c^{*T}$.
}
\newcolumntype{L}{>{$}l<{$}}
\newcolumntype{R}{>{$}r<{$}}
\newcolumntype{C}{>{$}c<{$}}
\begin{ruledtabular}
\begin{tabular}{CCCCCCCC}
 z & B_c & B_c^* & B_c^{*T} & z & B_c & B_c^* & B_c^{*T}  \\
\hline
 0.05 & 0.2067948 & 1.4072295 & 0.9439353 & 0.55 & 0.1088787 & 0.14250261 & 0.0995547 \\
 0.10 & 0.2324091 & 1.1553789 & 0.7809509 & 0.60 & 0.0900531 & 0.11125908 & 0.0777975 \\
 0.15 & 0.2370119 & 0.9379125 & 0.6375713 & 0.65 & 0.0720573 & 0.08640575 & 0.0604093 \\
 0.20 & 0.2310801 & 0.7550208 & 0.5156830 & 0.70 & 0.0551871 & 0.06615581 & 0.0461942 \\
 0.25 & 0.2191640 & 0.6033209 & 0.4138126 & 0.75 & 0.0397558 & 0.04906748 & 0.0341859 \\
 0.30 & 0.2036852 & 0.4789394 & 0.3297836 & 0.80 & 0.0261184 & 0.03411082 & 0.0236944 \\
 0.35 & 0.1860753 & 0.3780786 & 0.2612879 & 0.85 & 0.0147157 & 0.02081929 & 0.0144107 \\
 0.40 & 0.1672448 & 0.2971634 & 0.2060682 & 0.90 & 0.0061426 & 0.00958561 & 0.0066094 \\
 0.45 & 0.1478140 & 0.2328895 & 0.1619915 & 0.95 & 0.0011627 & 0.00202440 & 0.0013901 \\
 0.50 & 0.1282375 & 0.1822416 & 0.1270849 & 0.99 & 0.0000139 & 0.00002650 & 0.0000181 \\
\end{tabular}
\end{ruledtabular}
\end{table}

\begin{figure}[tb]
\centering
\includegraphics[width=0.45\textwidth]{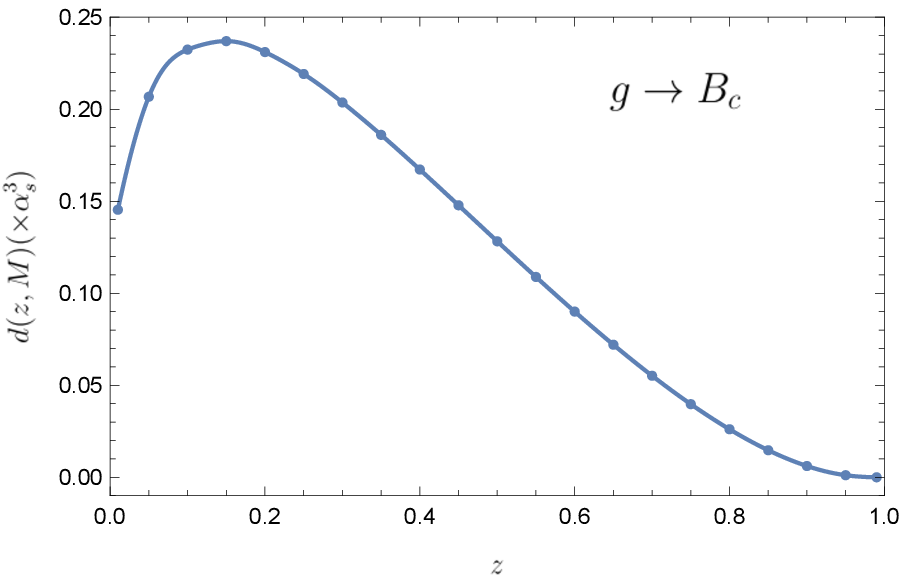}
\includegraphics[width=0.45\textwidth]{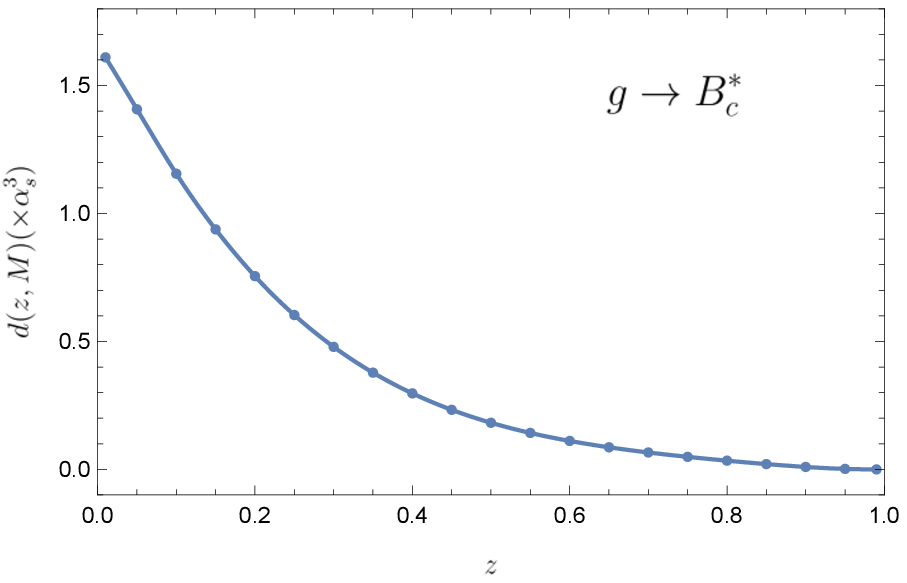}
\includegraphics[width=0.45\textwidth]{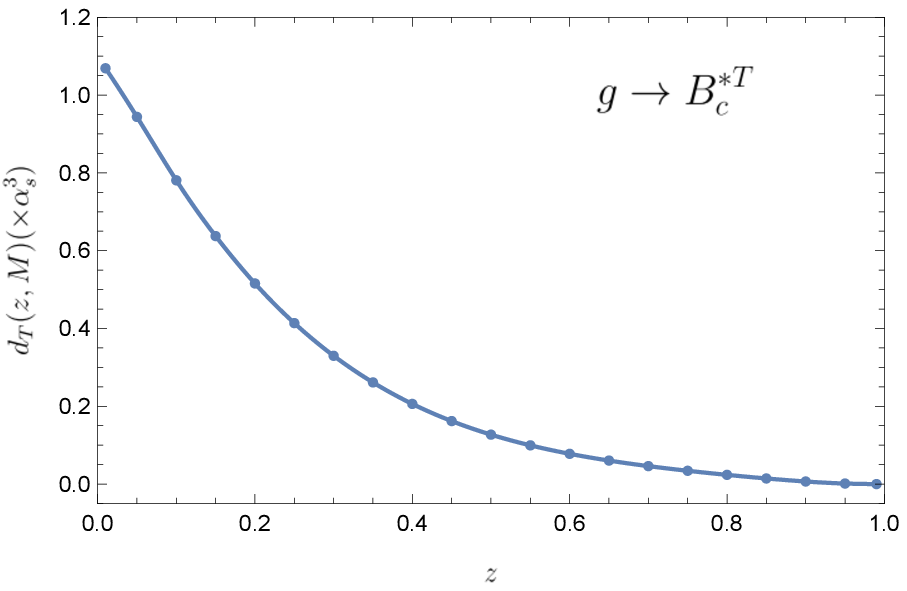}
\caption{Profiles of non-logarithmic coefficient functions
$c_1^H(z)$ defined in \eqref{Dividing:SDC:two:parts}.}
\label{Fig:2}
\end{figure}

For reader's convenience, in Table~{\ref{c1-table}} we have enumerated the values of $c^{H}_{1}(z)$
for some typical values of $z$. We also plot these functions in Fig.~\ref{Fig:2}.

\begin{figure}[tb]
\centering
\includegraphics[width=0.45\textwidth]{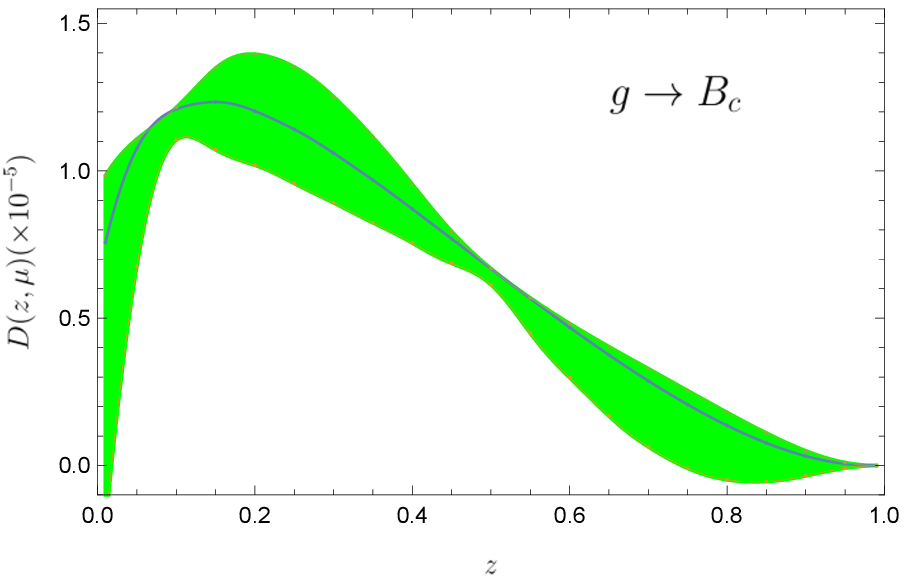}
\includegraphics[width=0.45\textwidth]{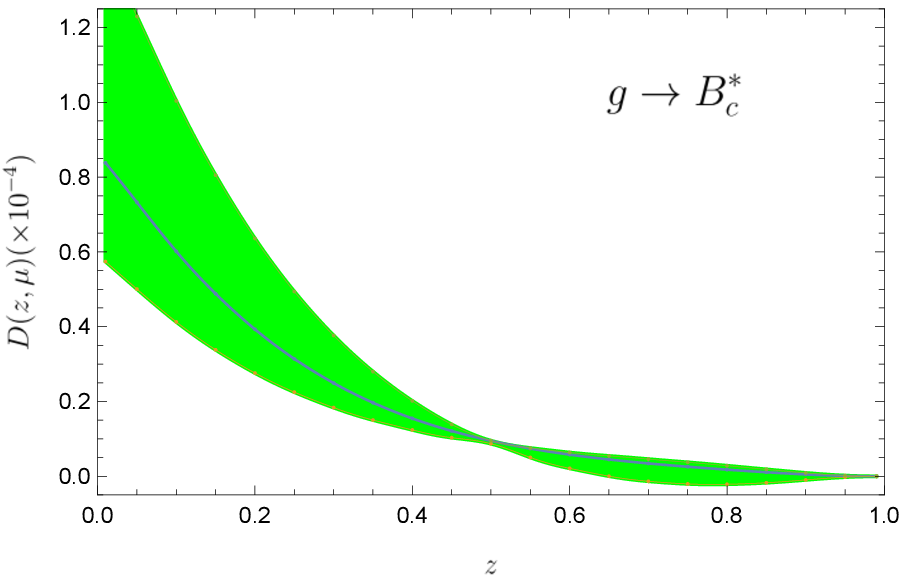}
\includegraphics[width=0.45\textwidth]{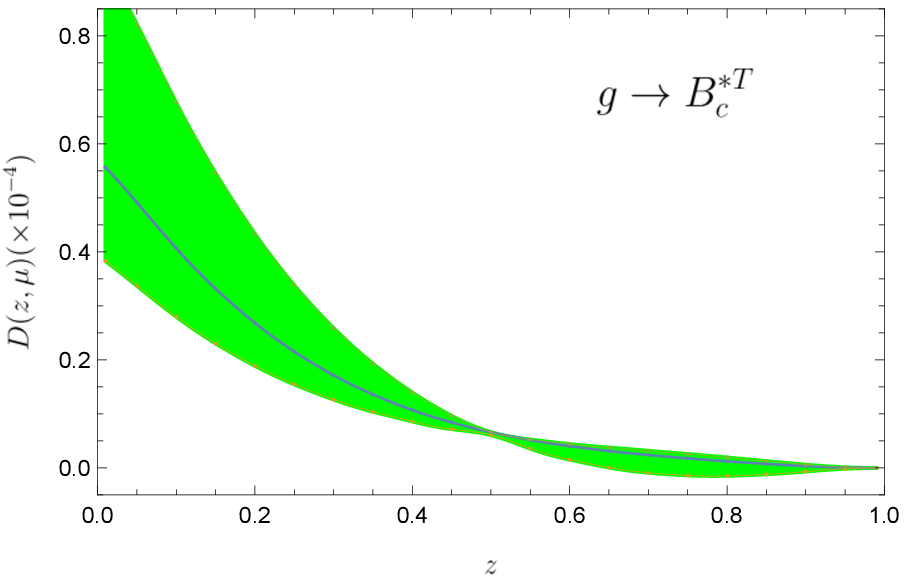}
\caption{The profiles of the fragmentation functions $D_{g\to H}(z,\mu)$ with $m_c=1.5\:{\rm GeV}$ and $m_b=4.8\:{\rm GeV}$, and
$\vert R(0)\vert^2 = 1.642\:{\rm GeV}^3$.
The uncertain band is obtained by sliding $\mu$ from $M/2$ to $2M$, with the central line taken at $\mu=M=m_b+m_c$.}
\label{Fig:3}
\end{figure}

In Fig.~\ref{Fig:3}, we also plot the various fragmentation functions $D_{g\to H}(z)$ for $H=B_c,B_c^*$ and $B_c^{*T}$ as function of $z$.
We observe that these functions are regular near $z=0$.

In summary, in this work we have computed the $g$-to-$B_c^{(*)}$ fragmentation functions within
NRQCD factorization framework, at the lowest order in velocity expansion and $\alpha_s$.
We start from the Collins and Soper's gauge-invariant operator definition, which makes the
renormalization program transparent.  We have employed an automated approach based on
sector decomposition strategy to conduct the phase space integral in dimensional regularization.
It turns out that, within tolerable amount of time, this method yields better numerical accuracy than the conventional NLO subtraction method.
After implementing DGLAP renormalization procedure, we obtain the UV finite
NRQCD short-distance coefficient functions in a pointwise manner.
We hope that our results are helpful to strengthen our understanding of
$B_c^{(*)}$ production at large-$p_T$ at \texttt{LHC} experiment.

\vspace{0.5cm}
\noindent{\it Note added.} After this work was finished and while we were preparing the manuscript,
a preprint has appeared very recently~\cite{Zheng:2021sdo}, which also computes the gluon-to-$B_c^{(*)}$ fragmentation
functions in NRQCD factorization approach, albeit using the conventional subtraction method.

\begin{acknowledgments}
We thank Wen-Long Sang for valuable discussions.
The work of F.~F. is supported by the National Natural Science Foundation of China under Grant No. 11875318 and by the Yue Qi Young Scholar Project in CUMTB.
 The work of  Y.~J. is supported in part by the National Natural Science Foundation of China under Grants No.~11925506, 11875263, No.~11621131001 (CRC110 by DFG and NSFC).
The work of D.~Y. is supported in part by the National Natural Science Foundation of China under Grants No.~11635009.
\end{acknowledgments}


\begin{thebibliography}{99}

\bibitem{Collins:1989gx}
  J.~C.~Collins, D.~E.~Soper and G.~F.~Sterman,
  Adv.\ Ser.\ Direct.\ High Energy Phys.\  {\bf 5}, 1 (1989)
  [hep-ph/0409313].

\bibitem{Braaten:1993mp}
  E.~Braaten, K.~m.~Cheung and T.~C.~Yuan,
  Phys.\ Rev.\ D {\bf 48}, 4230 (1993)
  [hep-ph/9302307].

\bibitem{Braaten:1993rw}
  E.~Braaten and T.~C.~Yuan,
  Phys.\ Rev.\ Lett.\  {\bf 71}, 1673 (1993)
  [hep-ph/9303205].

\bibitem{Bodwin:1994jh}
  G.~T.~Bodwin, E.~Braaten and G.~P.~Lepage,
  Phys.\ Rev.\ D {\bf 51}, 1125 (1995)
  Erratum: [Phys.\ Rev.\ D {\bf 55}, 5853 (1997)]
  [hep-ph/9407339].

\bibitem{Braaten:1993jn}
  E.~Braaten, K.~m.~Cheung and T.~C.~Yuan,
  Phys.\ Rev.\ D {\bf 48}, no. 11, R5049 (1993)
  [hep-ph/9305206].

\bibitem{Ma:1994zt}
  J.~P.~Ma,
  Phys.\ Lett.\ B {\bf 332}, 398 (1994)
  [hep-ph/9401249].

\bibitem{Braaten:1994kd}
  E.~Braaten and T.~C.~Yuan,
  Phys.\ Rev.\ D {\bf 50}, 3176 (1994)
  [hep-ph/9403401].

\bibitem{Cho:1994qp}
  P.~L.~Cho and M.~B.~Wise,
  Phys.\ Rev.\ D {\bf 51}, 3352 (1995)
  [hep-ph/9410214].

\bibitem{Ma:1995ci}
  J.~P.~Ma,
  Nucl.\ Phys.\ B {\bf 447}, 405 (1995)
  [hep-ph/9503346].

\bibitem{Ma:1995vi}
  J.~P.~Ma,
  Phys.\ Rev.\ D {\bf 53}, 1185 (1996)
  [hep-ph/9504263].

\bibitem{Braaten:1995cj}
  E.~Braaten and T.~C.~Yuan,
  Phys.\ Rev.\ D {\bf 52}, 6627 (1995)
  [hep-ph/9507398].

\bibitem{Cheung:1995ir}
  K.~m.~Cheung and T.~C.~Yuan,
  Phys.\ Rev.\ D {\bf 53}, 3591 (1996)
  [hep-ph/9510208].

\bibitem{Qiao:1997wb}
  C.~f.~Qiao, F.~Yuan and K.~T.~Chao,
  Phys.\ Rev.\ D {\bf 55}, 5437 (1997)
  [hep-ph/9701249].

\bibitem{Braaten:2000pc}
  E.~Braaten and J.~Lee,
  Nucl.\ Phys.\ B {\bf 586}, 427 (2000)
  [hep-ph/0004228].

\bibitem{Bodwin:2003wh}
  G.~T.~Bodwin and J.~Lee,
  Phys.\ Rev.\ D {\bf 69}, 054003 (2004)
  [hep-ph/0308016].

\bibitem{Sang:2009zz}
  W.~l.~Sang, L.~f.~Yang and Y.~q.~Chen,
  Phys.\ Rev.\ D {\bf 80}, 014013 (2009).

\bibitem{Bodwin:2012xc}
  G.~T.~Bodwin, U.~R.~Kim and J.~Lee,
  JHEP {\bf 1211}, 020 (2012)
  [arXiv:1208.5301 [hep-ph]].

\bibitem{Ma:2013yla}
  Y.~Q.~Ma, J.~W.~Qiu and H.~Zhang,
  Phys.\ Rev.\ D {\bf 89}, no. 9, 094029 (2014)
  [arXiv:1311.7078 [hep-ph]].

\bibitem{Artoisenet:2014lpa}
  P.~Artoisenet and E.~Braaten,
  JHEP {\bf 1504}, 121 (2015)
  [arXiv:1412.3834 [hep-ph]].

\bibitem{Bodwin:2014bia}
  G.~T.~Bodwin, H.~S.~Chung, U.~R.~Kim and J.~Lee,
  Phys.\ Rev.\ D {\bf 91}, no. 7, 074013 (2015)
  [arXiv:1412.7106 [hep-ph]].

\bibitem{Gao:2016ihc}
  X.~Gao, Y.~Jia, L.~Li and X.~Xiong,
  Chin.\ Phys.\ C {\bf 41}, no. 2, 023103 (2017)
  [arXiv:1606.07455 [hep-ph]].

\bibitem{Sepahvand:2017gup}
  R.~Sepahvand and S.~Dadfar,
  Phys.\ Rev.\ D {\bf 95}, no. 3, 034012 (2017).

\bibitem{Zhang:2017xoj}
  P.~Zhang, Y.~Q.~Ma, Q.~Chen and K.~T.~Chao,
  Phys.\ Rev.\ D {\bf 96}, no. 9, 094016 (2017)
  [arXiv:1708.01129 [hep-ph]].

\bibitem{Feng:2017cjk}
F.~Feng, S.~Ishaq, Y.~Jia and J.~Y.~Zhang,
Phys. Rev. D \textbf{102}, no.1, 014038 (2020)
[arXiv:1712.09986 [hep-ph]].

\bibitem{Yang:2019gga}
D.~Yang and W.~Zhang,
Chin. Phys. C \textbf{43} (2019) no.8, 083101
doi:10.1088/1674-1137/43/8/083101
[arXiv:1905.02923 [hep-ph]].

\bibitem{Artoisenet:2018dbs}
P.~Artoisenet and E.~Braaten,
JHEP \textbf{01}, 227 (2019)
doi:10.1007/JHEP01(2019)227
[arXiv:1810.02448 [hep-ph]].

\bibitem{Feng:2021uct}
F.~Feng, Y.~Jia and W.~L.~Sang,
Eur. Phys. J. C \textbf{81}, no.7, 597 (2021)
doi:10.1140/epjc/s10052-021-09390-4
[arXiv:2106.02027 [hep-ph]].

\bibitem{Zhang:2018mlo}
P.~Zhang, C.~Y.~Wang, X.~Liu, Y.~Q.~Ma, C.~Meng and K.~T.~Chao,
JHEP \textbf{04}, 116 (2019)
doi:10.1007/JHEP04(2019)116
[arXiv:1810.07656 [hep-ph]].

\bibitem{Feng:2018ulg}
F.~Feng and Y.~Jia,
[arXiv:1810.04138 [hep-ph]].

\bibitem{Collins:1981uw}
  J.~C.~Collins and D.~E.~Soper,
  Nucl.\ Phys.\ B {\bf 194}, 445 (1982).

\bibitem{Nogueira:1991ex}
  P.~Nogueira,
  J.\ Comput.\ Phys.\  {\bf 105}, 279 (1993).

\bibitem{Petrelli:1997ge}
  A.~Petrelli, M.~Cacciari, M.~Greco, F.~Maltoni and M.~L.~Mangano,
  Nucl.\ Phys.\ B {\bf 514}, 245 (1998)
  [hep-ph/9707223].

\bibitem{Mertig:1990an}
  R.~Mertig, M.~Bohm and A.~Denner,
  Comput.\ Phys.\ Commun.\  {\bf 64}, 345 (1991).

\bibitem{Feng:2012tk}
  F.~Feng and R.~Mertig,
  arXiv:1212.3522 [hep-ph].

\bibitem{Feng:2012iq}
  F.~Feng,
  Comput.\ Phys.\ Commun.\  {\bf 183}, 2158 (2012)
  [arXiv:1204.2314 [hep-ph]].

\bibitem{Binoth:2000ps}
  T.~Binoth and G.~Heinrich,
  Nucl.\ Phys.\ B {\bf 585}, 741 (2000)
  [hep-ph/0004013].

\bibitem{Binoth:2003ak}
  T.~Binoth and G.~Heinrich,
  Nucl.\ Phys.\ B {\bf 680}, 375 (2004)
  doi:10.1016/j.nuclphysb.2003.12.023
  [hep-ph/0305234].

\bibitem{Zheng:2019gnb}
X.~C.~Zheng, C.~H.~Chang, T.~F.~Feng and X.~G.~Wu,
Phys. Rev. D \textbf{100}, no.3, 034004 (2019)
doi:10.1103/PhysRevD.100.034004
[arXiv:1901.03477 [hep-ph]].

\bibitem{Zheng:2019dfk}
X.~C.~Zheng, C.~H.~Chang and X.~G.~Wu,
Phys. Rev. D \textbf{100}, no.1, 014005 (2019)
doi:10.1103/PhysRevD.100.014005
[arXiv:1905.09171 [hep-ph]].

\bibitem{Smirnov:2013eza}
  A.~V.~Smirnov,
  Comput.\ Phys.\ Commun.\  {\bf 185}, 2090 (2014)
  doi:10.1016/j.cpc.2014.03.015
  [arXiv:1312.3186 [hep-ph]].

\bibitem{Eichten:1995ch}
E.~J.~Eichten and C.~Quigg,
Phys. Rev. D \textbf{52}, 1726-1728 (1995)
doi:10.1103/PhysRevD.52.1726
[arXiv:hep-ph/9503356 [hep-ph]].

\bibitem{Zheng:2021sdo}
X.~C.~Zheng, C.~H.~Chang and X.~G.~Wu,
[arXiv:2112.10520 [hep-ph]].


\end{thebibliography}
\end{document}